\documentstyle[aps,prl,twocolumn,eqsecnum,floats,axodraw,psfig]{revtex}
\def\KeyWord#1{$\backslash$\IfColor{$\!\!$\textRed{#1}\textBlack}{#1}$\!\!$}

\begin{document}
\wideabs{

\title{
{ \hfill \small UCB-PTH-02/37  \\
\hfill  \small LBNL-51431 \\
 \hfill \small FERMILAB-Pub-02/300  \\
\hfill \small hep-ph/0208276 \\
} $~$\\ Supergravity Inflation Free from Harmful Relics }
\author{Patrick~B.~Greene$^{(1)}$, Kenji~Kadota$^{(2)}$ and 
Hitoshi~Murayama$^{(2)(3)}$ }

\address{$^{(1)}${\it NASA/Fermilab Astrophysics Group, Fermi National
        Accelerator Laboratory, Batavia, IL 60510, USA}}
\address{$^{(2)}${\it
        Department of Physics, University of California, Berkeley, CA
94720,
        USA}} \address{$^{(3)}${\it Theory Group, Lawrence Berkeley
National
        Laboratory, Berkeley, CA 94720, USA}}

\maketitle
\begin{abstract}

We present a realistic supergravity inflation model which is free
from the overproduction of potentially dangerous relics in
cosmology, namely moduli and gravitinos which can lead to the
inconsistencies with the predictions of baryon asymmetry and
nucleosynthesis.  The radiative correction turns out to play a
crucial role in our analysis which raises the mass of
supersymmetry breaking field to intermediate scale. We pay a
particular attention to the non-thermal production of gravitinos
using the non-minimal K\"ahler potential we obtained from loop
correction.  This non-thermal gravitino production however is
diminished because of the relatively small scale of inflaton mass
and small amplitudes of hidden sector fields.
\end{abstract}
}

\narrowtext

 \setcounter{footnote}{0} 
\setcounter{page}{1}
\setcounter{section}{0} \setcounter{subsection}{0}
\setcounter{subsubsection}{0}

\section{Introduction}

\setcounter{footnote}{0}

There exist generic and well known problems in constructing
supergravity inflation models with broken local supersymmetry in 
vacuum. First, one must carefully choose the superpotential
and K\"ahler potential such that non-renormalizable terms do not
spoil the flatness of the inflaton potential, which is also
essential to obtain the observed CMB spectrum. Second, most
supergravity inflation models run into cosmological problems at
late times due to an over-abundance of harmful relics such as
moduli and gravitinos,
and we need to check that a model is free from these problems 
to be consistent with the data of baryon asymmetry and
nucleosynthesis predictions.



Several supergravity inflation models free from abundant moduli
have been proposed \cite{holman,sarkar2,dine,izumi,sarkar}, and we
give a brief review for this moduli problem with emphasis on the
importance of the radiative corrections. Especially when the 
supersymmetry breaking field has flat direction at tree
level, radiative correction has a significant effect on its
potential to lead to the necessity for the modification of minimal
K\"ahler potential.

More recently, the nonthermal production of gravitinos has been
drawn attention \cite{long,marco,toni,dangerous,firstgrav}. These
problems on gravitino production during preheating era, however,
have been analyzed only using the minimal K\"ahler potential so
far. The gravitino interactions depend on the form of K\"ahler
potential and consequently the non-thermal production of
gravitinos can depend on its form as well. Our treatment is the
first analysis of gravitino non-thermal production taking 
account of the non-minimal K\"ahler potential obtained from the
loop correction of supersymmetry breaking field.




The detailed analysis for nonthermal production of gravitinos in a
system of coupled fields has been done only for models with
non-renormalizable hidden sector supersymmetry breaking, and only
the Polonyi model in particular\cite{marco,marco2}. It was shown,
in this model, that dominant fermion fields which are created
efficiently through preheating mechanism are inflatinos rather
than gravitinos (thus free from gravitino problem). This type of
supergravity inflation model is a good toy model to investigate
the nonthermal particle production in preheating era, but is not a
realistic inflation model in that it
 sill suffers from Polonyi problem which is as serious a
problem as gravitino problem.  We give the first realistic
supergravity inflation model in this sense where nonthermal
production of gravitinos as well as moduli problem were explicitly
analyzed.

The paper is structured as follows.  In section \ref{setup}, we
explain our choice of superpotential and discuss how inflation
develops in our model. We then discuss how supersymmetry breaking
field evolves and calculate its radiative correction and its
modification to minimal K\"ahler potential arising from this loop
correction.  In section \ref{dangerousrelics}, we see if our model
leads to any cosmological crisis, namely, moduli and gravitino
problems. We give the conclusion and discussion at the end.

%
%
%

\section{Setup}
\label{setup}

We consider the superpotential \cite{holman} consisting of the
inflaton sector and hidden sector of O'Raifeartaigh type \cite{or}
which are gravitationally coupled to each other.
\begin{equation}
W=\Delta^2~
\frac{(\Sigma-M)^2}{M_p}+\Phi_1\left(\kappa\Phi_2^2-\mu^2\right)+\lambda\Phi_2\Phi_3+C,
\label{sup}
\end{equation}
where superfield $\Sigma$ includes inflaton scalar component
$\sigma$ and $\Phi_1$ includes O'Raifeartaigh scalar field
$\phi_1$. $ \Delta \sim 10^{-4} M_p$ from COBE normalization and
$M$ is set to $M_p (= 2.436 \cdot 10^{18} \mathrm{GeV} ) $ so that
the inflaton potential keeps the flatness around the origin (i.e.
for $ \frac{\partial}{\partial \sigma}V(0) \simeq 0,
\frac{\partial^2}{\partial \sigma^2}V(0) \simeq 0 $ ).  The
dimensionless parameter $\kappa$ is of order unity while the other
mass parameters $\mu$ and $\lambda$ are of intermediate scale 
( $\sim 10^{-8} M_p $ ).  $C$ is the constant term to cancel the
cosmological constant at the vacuum. We start with the discussion
for the evolution of scalar fields in inflaton sector and hidden
sector which are only gravitationally coupled to each other and,
for the sake of clarity, we first treat each sector separately
followed by the discussion including the coupling with non-minimal
K\"ahler potential arising from the radiative correction. We
assume gauge singlets in the potential for simplicity in the
following.
\subsection{Inflaton Sector}
The superpotential of inflaton sector is given as
\begin{equation}
W_{inflaton}=\Delta^2~ \frac{(\Sigma-M_p)^2}{M_p}.
\label{sup}
\end{equation}
The general expression for supergravity potential becomes, for
K\"ahler potential, $K$, and superpotential, $W$,
\begin{equation}
V = m_i~{K^{-1}}_j{}^i ~m^j- 3~M_p^{-2} \vert m \vert^2
\label{SUGRApot}
\end{equation}
with
\begin{eqnarray}
K^i{}_j\equiv \frac{\partial^2 K }{\partial \phi_i
\partial\phi^j},~~ m \equiv \mathrm{e}^{\frac{K}{2M_p^2}} W \\ m^i
\equiv D^i m \equiv \partial^i m + \frac{1}{2 M_p^2} ~(\partial^i
K) ~m ~. \label{eqn:m}
\end{eqnarray}
If we consider, for the moment, the minimal form of K\"ahler
potential,
$K=\Sigma^\dagger\,\Sigma$, the effective supergravity potential
from $W_{inflaton}$ for the real part of scalar component,
$\sigma$, becomes ( in natural units )
\begin{eqnarray}
V_{inflaton}=\mathrm{e}^K \left(\left \vert \frac{\partial
W}{\partial \Sigma}+ \Sigma^{\dag}W \right\vert^2-3\vert W\vert^2
\right) \label{eq:minimal}
\\
=\Delta^{4}~\mathrm{e}^{\sigma^2/2}~\left(1-\frac{\sigma^2}{2}
-\sqrt{2}\sigma^3+\frac74\sigma^4-\frac{1}{\sqrt{2}}\sigma^5+\frac{\sigma^6}{8}\right).
\label{eqn:Vinflaton}
\end{eqnarray}
We here point out the absence of linear and quadratic terms which
enables the potential to keep the flatness around the origin. We
can obtain the value of $\Delta\sim10^{-4} M_p$
 from COBE normalization condition\cite{white} \footnote{ \cite{holman}
gives an order of estimates $10^{-4}M_p\leq \Delta \leq
10^{-3.5}M_p$ from the constraints on gravitino abundance and
proton decay.},
\begin{equation}
\left(\frac{V}{\epsilon}\right)^{\frac{1}{4}}\simeq0.027M_p (
1-3.2 \epsilon + 0.5 \eta),
\end{equation}
which  should be evaluated at the horizon exit. Scale of inflaton
field when the inflation ends and the cosmological scales leave
the horizon are obtained from slow-roll conditions, $\epsilon
\lesssim 1$, $\eta \lesssim 1$, and 60 e-folding
 condition,
\begin{equation}
N(\sigma_{exit}) \simeq
\int\limits_{\sigma_{end}}^{\sigma_{exit}}\frac{V}{V'}d\sigma \,
\simeq 60.
\end{equation}
  We also note the scale of inflation is of the
order $\Delta^{4}$ and the mass of the inflaton is of the order
$\Delta^{2}/M_p $ with its decay width $ \Gamma_{\sigma}\simeq
m_{\sigma}^3/M_p^2=\Delta^6/M_p^5 $ assuming gravitational
strength coupling to ordinary fields.
\subsection{Hidden Sector}
The supersymmetry breaking sector is that of O'Raifeartaigh model,
\begin{equation}
W_{hidden}=\Phi_1\left(\kappa\Phi_2^2-\mu^2\right)+\lambda\Phi_2\Phi_3+C
~~.
\end{equation}

This is a familiar example of supersymmetry breaking due to
non-vanishing
 $F$-term from $\Phi_1$, $\vert F \vert=\mu^2 $, in the vacuum.
Therefore we
 add $C=\frac{\mu^2}{\sqrt{3}}M_p$ (compare with $\frac{-3\vert
W\vert^2}{M_P}$
 term in eqn(\ref{eqn:Vinflaton}) ) for the vanishing cosmological
constant at
 the vacuum.\footnote{ This additional constant term $C$, strictly
speaking,
 should be modified if we include the radiative correction and the
coupling
 between inflaton and hidden sectors.  We, however, stick to this value
of $C$
 for simplicity because this modification essentially does not change
our
 discussion.} We should, however, expect that, when the fields are far
away
 from the vacuum during the inflation, there are additional $F$-terms
from
 other fields in the effective scalar potential and these $F$-terms
lead to
 additional `cosmological constant' $ \Lambda^{4}$ \cite{dine}. Adding
 ${e}^{K}\Lambda^{4}$ in the potential indicates us that this
cosmological
 constant term during the inflation gives an additional effective mass
of order
 $\frac{\Lambda^{4}}{M_p^{2}}$ to each field in the model. Fields
$\phi_2$ and
 $\phi_3$ do not posses the linear terms and these two fields quickly
roll down
 to the origin ( i.e. to their minimum) during the inflation. The
scalar field
 $\phi_1$, however, has a liner term and its minimum shifts according
to the
 evolution of inflaton field as we shall see in the next section.
Because we
 are interested in the particle production after the inflation, we
focus on the
 evolution of O'Raifeartaigh field $\phi_1$ among the fields in this
 supersymmetry breaking sector.  Moreover, the $F$-term at the vacuum
has the
 contribution only from the scalar field $\phi_1$ which turns out to
have flat
 direction at the tree level. Because of this flat potential with
respect to
 $\phi_1$ at the tree level, the global supersymmetry radiative
corrections
 have a significant effect on the effective potential and consequently
give
 non-negligible modification to the minimal K\"ahler potential. The
radiative
 corrections of local supersymmetry are always Planck mass suppressed
and we do
 not consider them here. We calculated this non-minimal
K\"ahler
 potential from the calculation of loop correction \cite{huq,jac}.
Setting the
 parameter range to be $2\kappa\mu^2 < \lambda^2$ to make $ \phi_2$ and
$\phi_3
 $ stay at the origin in the vacuum leads to the following one-loop
correction
 due to $\phi_1$,
\begin{eqnarray}
V_{one\,loop} = \frac{1}{ 64\pi^{2} } \left( \sum_{i=1}^{4}
(M_{i}^{2})^2 \left[ \log\left(\frac{M_{i}^{2}}{\lambda^{2}}\right) -
\frac{3}{2}\right]  \right.\\
\left. -2\sum_{i=1}^{2}(N_{i}^{2})^{2} \left[
\log\left(\frac{N_{i}^{2}}{\lambda^{2}}\right) - \frac{3}{2}\right] \right),
\nonumber \\
\end{eqnarray}
where we have defined
\begin{eqnarray}
& &M_{1}^2=\frac{1}{2}(A_{1}-A_{2}),
M_{2}^2=\frac{1}{2}(A_{1}+A_{2}), \nonumber \\ &
&M_{3}^2=\frac{1}{2}(A_{3}-A_{4}),
M_{4}^2=\frac{1}{2}(A_{3}+A_{4}), \nonumber
\\ & &N_1^2=\frac{1}{2}(B_1-B_2), N_2^2=\frac{1}{2}(B_1+B_2), \nonumber
\\ &
&A_1=2\lambda^{2}-2\kappa\mu^{2}+4\kappa^{2}\vert \phi_1 \vert
^{2},~~ A_3=2\lambda^{2}+2\kappa\mu^{2}+4\kappa^{2}\vert
\phi_1\vert ^{2},
\nonumber \\
& &A_2=\sqrt{ 4\kappa^{2}\mu^{4}+16\lambda^{2}\kappa^{2}\vert
\phi_1 \vert ^{2}- 16\kappa^{3}\mu^{2}\vert \phi_1 \vert
^{2}+16\kappa^{4} \vert \phi_1 \vert ^{4} }, \nonumber \\ &
&A_4=\sqrt{ 4\kappa^{2}\mu^{4}+16\lambda^{2}\kappa^{2} \vert
\phi_1 \vert ^{2} +16\kappa^{3}\mu^{2} \vert \phi_1 \vert
^{2}+16\kappa^{4} \vert \phi_1 \vert ^{4} }, \nonumber \\ &
&B_1=2\lambda^{2}+4\kappa^{2} \vert \phi_1 \vert ^{2},~~
B_2=\sqrt{ 16\lambda^{2}\kappa^{2} \vert \phi_1 \vert ^{2}
+16\kappa^{4} \vert
\phi_1 \vert ^{4} }\nonumber \, . \\
\end{eqnarray}
We have used the $\overline{MS}$ scheme and taken the
renormalization scale to be $\lambda$. We are concerned with the
regime $|\phi_1| \lesssim \lambda/\kappa$, as we shall show in the
next section.  In this case, and for $2\kappa\mu^2 \ll \lambda^2$,
we can approximate the one loop potential as
\begin{equation}
V_{one\,loop} = C_1 + \frac{\kappa^{2}\mu^{4}}{8\pi^{2}}
 \left(\frac{\kappa^{2}\vert \phi_1 \vert
 ^{2}}{\lambda^2} + \ldots \right) \, ,
\label{Vloop}
\end{equation}
 where $\ldots$ are terms of order
$\kappa^4|\phi_1|^4/\lambda^4$ and higher and $C_1$ is a
small constant term ($\ll \mu^4$) which can be absorbed into the
constant part of the superpotential. We can find the following
one loop
correction to the K\"ahler potential
($K=K_{minimal}+K_{correction}$), by comparing eqn(\ref{Vloop})
with eqn(\ref{SUGRApot}), 
\begin{equation}
K_{correction}=- \frac{\kappa^{2}}{32\pi^{2}}
\left(\frac{\kappa^{2}\Phi_1^2 \Phi_1^{\dagger 2} }{\lambda^2}
\right) . \label{eqn:correction}
\end{equation}
We shall use this non-minimal K\"ahler potential in our analysis
for
 non-thermal production of gravitinos.  We note that this radiative
 correction enhances the coupling to the longitudinal component of
gravitino
 and raises the mass of $\phi_1$ which was massless at the tree level
to the
 intermediate scale
\begin{equation}
m_{\phi_1}^2= \frac{\kappa^4\mu^4}{4\pi^2\lambda^2}= \frac{
\alpha_{\kappa}^2\mu^4}{ \lambda^2} ~\mbox{with} ~ \alpha_{\kappa}
\equiv \frac{\kappa^2}{4\pi}, \label{eqn:ormass}
\end{equation}
which turns out to be crucial to evade the moduli problem.

\section{Harmful Relics}
\label{dangerousrelics}

We are now in a position to discuss the fate of inflaton and
O'Raifeartaigh fields in the coupled effective potential with
non-minimal K\"ahler potential to see if our model leads to any
cosmological crisis.  We first briefly review the resolutions of
so-called Polonyi or moduli problem and we further discuss
the non-thermal production of gravitinos. \\
\subsection{ Moduli Problem}
We here start with the discussion on well-known potentially
dangerous problems, Polonyi problem or, in general, moduli
problem.  There are two aspects which we should worry about before
our discussion on decay of moduli into gravitinos. One is the case
when the moduli decay very late ( i.e. during or after the
nucleosynthesis) which can jeopardize nucleosynthesis predictions
because of ultra-relativistic decay products directly from moduli
fields destroying the light elements, in particular, ${}^4$He and
D nuclei. The other is when the entropy release due to its decay
is so big that it can over-dilute the baryon asymmetry well below
its acceptable amounts (so-called `entropy crisis').  Our model
does not have either of these problems because the radiative
correction raises its mass to as much as intermediate scale. Its
decay width is indeed enhanced up to $\Gamma_{\phi_1}\simeq
m_{\phi_1}^5/\vert F \vert^2\simeq \frac{\left(\mu^2
\alpha_{\kappa}/ \lambda\right)^5}{\mu^4} \simeq
\alpha_{\kappa}^5\mu $ with $ \alpha_{\kappa} \equiv \kappa^2/4\pi
\simeq O(10^{-1})$ and this is of order$10^{-13} M_p$.  So
O'Raifeartaigh field decays around $ 10^{13} M_p^{-1}$ in our
model which is much before the nucleosynthesis starts around $\sim
10^{40}M_p^{-1} $ and even well before the reheating starts due to
the inflaton decay around $ 1/\Gamma_{\sigma}\sim M_p^5/\Delta^6
\sim 10^{25} M_p^{-1} $.  We however need a great care about the
possibility of decay products with long life-time, especially
gravitinos. Gravitino decay rate is of order $ \Gamma_{m_{3/2}}
\sim m_{3/2}^3/M_p^2 \sim \mu^6/M_p^5 \sim 10^{-48}M_p $ and its
relativistic decay products, especially ultra-relativistic
photon/photino, can destroy the light elements in nucleosynthesis
(photo-dissociation process) as we just mentioned. The possible
abundant gravitino production from O'Raifeartaigh fields can be
caused by the energy release stored during the inflation by the
shift of the minimum of SUSY breaking field potential as inflaton
evolves.  If this energy release is too big, it could lead to
large amount of gravitinos and upset the nucleosynthesis
predictions.  We can see this is not the case for our model as
follows \cite{holman,izumi,randall}.  During the inflation, due to
the coupling to the inflaton ($\sigma \sim M_p $), O'Raifeartaigh
field amplitude is around the intermediate scale of order $\phi_1
\simeq \frac{\mu^2}{\Delta^2}M_p $ at the minimum of its
potential. Therefore we can estimate the energy stored in this
O'Raifeartaigh field when it starts oscillation ( i.e.
$t_{\phi_1}\simeq m_{\phi_1}^{-1}$) to be at most of order
\begin{equation}
\rho(t_{\phi_1}) \simeq m_{\phi_1}^2 \frac{\mu^4}{\Delta^4}M_p^2
\end{equation}
and its number density $n_{\phi_1}$ in this coherently oscillating
O'Raifeartaigh field is at most
\begin{equation}
n_{\phi_1}(t_{\phi_1})\simeq m_{\phi_1}
\frac{\mu^4}{\Delta^4}M_p^2 .
\end{equation}
We can now estimate its number density to entropy ratio at the
time of
 reheating for the gravitinos through the decay of $\phi_1$ ( at $
t=t_r $, say ) in an adiabatically expanding universe.  Assuming, for the upper
bound,
 $\phi_1$ solely decays into gravitinos with 100$\%$ branching ratio
and using
$ s \sim \frac{2 \pi^2}{45} g_* T^3 $ (with $g_*$ effective degree
of freedom) and $a^3\sim t^2$ for matter domination \cite{kolb}
due to the coherently oscillating inflaton field which dominates
the energy in the universe,
\begin{eqnarray}
\frac{n_{3/2} (t_r)}{s(t_r)}
\simeq\frac{n_{\phi_1}(t_{\phi_1})\left(
\frac{a(t_{\phi_1})}{a(t_r)} \right )^3}{0.44g_* T_{RH}^3} \nonumber \\
\simeq \frac{n_{\phi_1}(t_{\phi_1})\left( \frac{t_{\phi_1}} {t_r}
\right )^2}{0.44 g_* T_{RH}^3} .
\end{eqnarray}
This can lead to the estimate of $n_{3/2}/s$ after reheating by
substituting $1.66 \cdot T_{RH}^2 \sqrt{g_*} / M_p \sim H \sim
t_r^{-1}$ for $t_r$,
\begin{equation}
\frac{n_{3/2}}{s} \simeq \frac{n_{\phi_1}(t_{\phi_1}) t_{\phi_1}^2
T_{RH} ( 1.66)^2} {0.44 M_p^2} \simeq \frac{ (1.66)^2 \mu^4 T_{RH}
} { 0.44 m_{\phi_1} \Delta^4}.  \label{eqn:orns}
\end{equation}
We can now compare this value with one corresponding to the
gravitinos produced by the scattering in the thermal bath in
reheating era obtained in MSSM\cite{kawasaki,moroi2},
\begin{equation}
n/n_{rad}(T \ll 1\mbox{MeV}) \simeq 1.1 \cdot 10^{-11} \left (
\frac{T_{RH}}{10^{10} \mbox{GeV} } \right) .
\nonumber \\
\label{eqn:whatever}
\end{equation}
Using $s=1.8\cdot g_* n_{rad}$ and $g_*(\ll MeV)\simeq 3.36$, we
obtain
\begin{equation}
 n/s \simeq 1.8 \cdot 10^{-12} \left( \frac{T_{RH}}{10^{10} \mbox{GeV }
} \right) \label{eqn:ns} ~.
\end{equation}
flow of the gauge coupling.
We can now transform eqn(\ref{eqn:orns}) by substituting
(\ref{eqn:ormass}) to the following form,
\begin{equation}
n/s \simeq 3.5 \cdot
10^{-14}\left(\frac{T_{RH}}{10^{10}\mbox{GeV}}\right) ~~.
\end{equation}
 This is smaller than the thermal production of gravitino
(\ref{eqn:ns}) by two
 orders of magnitude. The radiative correction therefore induces
intermediate
 mass scale for O'Raifeartaigh field and it consequently makes the
 O'Raifeartaigh field energy released through the decay into gravitinos
after
 the inflation small enough to evade the abundant gravitinos.  Hence
our model
 does not suffer from moduli problem as far as the constraint from
thermal
 gravitino production is satisfied.

We mention that non-adiabatic production of moduli
 fields in pre-heating era could lead to
 abundant gravitinos\cite{premoduli}.
We are, however, not concerned about the
 parametric resonance effects for moduli fields
because scalar coupling terms in Lagrangian in our model are
trilinear in hidden sector fields, and those couplings to inflaton
field have always Planck mass suppression \cite{sarkar}.

The exception where this Planck mass suppression does not occur
and preheating effects could be important
 is the coupling involving longitudinal components of
gravitino, which is the subject in the following section.

\subsection{Non-thermal Production of Gravitino}
\label{sec:nonthermal} It has been argued recently that parametric
resonance mechanism in preheating era for the creation of
gravitino can be much more efficient than the thermal one
\cite{long,marco,toni,dangerous,firstgrav}.  In this
nonperturbative mechanism, the gravitinos can be created
non-adiabatically through the amplification of vacuum fluctuation
via rapid energy transfer from coherently oscillating inflaton
field which still dominates the energy density in the universe
just after inflation and before the reheating era.

In analyzing the gravitino field equations in the following, we
see that the equations for transverse and longitudinal components
of gravitino decouple. While transverse component equation has a
Planck mass suppressed coupling and thus gravitationally
suppressed particle creation, longitudinal component equation is
free from Planck mass suppression and it could lead to the
abundant gravitino production well above the constraint from
thermal production of gravitino.
 Indeed, gravitino-goldstino equivalence theorem states that the
equation for gravitino longitudinal component can be reduced to
the equation of goldstino in global supersymmetry in the limit of
weak gravitational coupling.  This warns us that gravitino
longitudinal components could lead to its efficient copious
production without Planck mass suppression.  Our model however has
a Planck scale amplitude for inflaton field after inflation, and
it is not obvious if this naive intuitive picture analogous to the
goldstinos in global SUSY applies here.
Therefore we apply here the formalism developed in \cite{long,marco} 
to calculate the number density of gravitinos created through the
nonthermal process.

We first need to describe the evolution of scalar fields and
fermion fields and their interactions.  It is convenient to work
with, among other possible choices, the following rescaled
quantities in our numerical analysis,
\begin{eqnarray}
\lefteqn{ {\hat \phi_1} \equiv\frac{\phi_1}{M_p}, ~~ {\hat \phi_2}
\equiv\frac{\phi_2}{M_p},~~ {\hat \phi_3}
\equiv\frac{\phi_3}{\Delta},~~ {\hat \sigma} \equiv
\frac{\sigma}{M_p},~~ {\hat \mu} \equiv\frac{\mu}{\Delta},
\nonumber }\\ & & {\hat \lambda} \equiv \frac{\lambda}{\Delta},~~
{\hat t} \equiv t~\frac{ \Delta^2}{M_P},~~ {\hat H} \equiv H
\frac{M_p}{\Delta^2},~~ {\hat V} \equiv \frac{V}{\Delta^4} ,
\end{eqnarray} where $H$ is Hubble constant, $V$ is a scalar
potential from eqn (\ref{SUGRApot}) with non-minimal K\"ahler
potential obtained in (\ref{eqn:correction}),
\begin{eqnarray}
\!\!\!K=\Sigma\Sigma^{\dag}+\Phi_1\Phi_1^{\dag}+\Phi_2\Phi_2^{\dag}+
\Phi_3\Phi_3^{\dag} \nonumber \\ - \frac{\kappa^{2}}{32\pi^{2}}
\left(\frac{\kappa^{2} \Phi_1^2 \Phi_1^{\dag2 }}{\lambda^2}
\right) .
\end{eqnarray}
 In terms of these rescaled quantities, the equations of motion for
coherently oscillating scalar fields $\phi$($= \sigma, \phi_i$)
read
\begin{equation}
\frac{d^2 {\hat \phi}}{d {\hat t}^2} + 3 \, {\hat H} \, \frac{d
{\hat \phi}}{d {\hat t}} + \frac{d {\hat V}}{d {\hat \phi}} = 0
~~.
\end{equation}
We omit $\hat{}$~ in the following discussion as long as it is
clear from the
 contexts.  We can concentrate on the field equations for $\sigma$ and
$\phi_1$
 because the other fields in supersymmetry breaking sector quickly roll
down to
 the origin during the inflaton and stay there\footnote{ Once these
fields roll
 down to the origin, they stay at the origin to any higher order
because of
 R-symmetry.}. So we can let the amplitudes of $\phi_2$ and $\phi_3$
vanish
 after obtaining the equation of motion for $\sigma$ and $\phi_1$ to
see the
 field evolutions after the inflation.


The Fermion equation follows from the supergravity Lagrangian
\begin{eqnarray}
e^{-1} L = -\frac{1}{2} \, M_p^2 \, R - K_i{}^j \left(
\partial_\mu \, \phi^i \right) \left( \partial^\mu \phi_j \right)
- V \nonumber\\ -\,\frac12 \, M_p^2 \, \bar \psi_\mu \, R^\mu +
\frac12 \, m \, \bar \psi_{\mu R} \, \gamma^{\mu \nu} \, \psi_{\nu
R} \nonumber\\ +\,\frac12 \, m^* \, \bar \psi_{\mu L} \,
\gamma^{\mu\nu } \, \psi_{\nu L} - K_i{}^j \left[ \bar \chi_j \,
\not\!\! D \bar \chi^i + \bar \chi^i \not\!\! D \bar \chi_j
\right] \nonumber \\ - m^{ij}
\, \bar \chi_i \, \chi_j - m_{ij} \, \bar \chi^i \, \chi^j \nonumber \\
+ \left( 2 \, K_j{}^i \bar \psi_{\mu R} \, \gamma^{\nu \mu} \,
\chi^j \,
\partial_\nu \phi_i + \bar \psi_R \cdot \gamma \upsilon_L + \mbox{h.c.}
\right) \nonumber \\ +\mbox{( four fermion and gauge interaction
terms )}. \label{lag}
\end{eqnarray}
This Lagrangian includes chiral complex multiplets $( \phi_i,\,
\chi_i )$ and the Ricci scalar $R$.  Subscript $L$ and $R$ denote
its projection through operators $P_L \equiv ( 1 + \gamma_5 )/2,
~P_R \equiv ( 1 - \gamma_5 )/2$. Gravitino kinetic term shows up
in the form of
\begin{equation}
R^\mu = e^{-1} \, \epsilon^{\mu \nu \rho \sigma} \, \gamma_5 \,
\gamma_\nu \, D_\rho \, \psi_\sigma\,,
\end{equation}
with covariant derivative
\begin{equation}
D_\mu \psi_\nu = \left( \left( \partial_\mu + \frac{1}{4}
\omega_\mu^{m n} \gamma_{m n} \right) \delta_\nu^\lambda -
\Gamma_{\mu \, \nu}^\lambda \right) \psi_\lambda\,.
\end{equation}
The kinetic term for chiral fermion is
\begin{eqnarray}
\lefteqn{D_\mu \chi_i \equiv \left( \partial_\mu + \frac{1}{4}
\omega_\mu^{m n} \gamma_{m n} \right) \chi_i \nonumber }\\ & &+
\frac{1}{4M_p^2} \left[
\partial_j K \partial_{\mu}\phi^j - \partial^j K \, \partial_\mu \phi_j
\right] \chi_i + \Gamma_i^{j\,k} \chi_j \partial_\mu \phi_k\,
\end{eqnarray}
with K\"ahler connection $\Gamma_i^{j\,k} \equiv {K^{-1}}_i{}^l
\partial^j
K_l{}^k $ and $\gamma_{m\,n} \equiv [ \gamma_m, \gamma_n ] /2$ .
Its mass term reads
\begin{equation}
m^{i j} \equiv D^i D^j m = \left( \partial^i + \frac{1}{2M_p^2}
(\partial^i K ) \right) m^j - \Gamma_k^{i\,j} \, m^k\,.
\end{equation}
The combination of matter fields gives left-handed component of
goldstino
\begin{equation}
\upsilon_L \equiv m^i \, \chi_i + \left( \not\!\partial \phi_i
\right) \chi^j \, K_j{}^i\,.\label{gold}
\end{equation}
The supersymmetry transformation of goldstino \cite{long}
\begin{equation}
\delta\upsilon=-\frac{3M_p^2}{2}(m_{3/2}^2 + H^2)~\epsilon,~
m_{3/2}\equiv\frac{\vert m \vert }{M_p^2}
\end{equation}
indicates that gravitino mass and Hubble parameter signal
supersymmetry breaking.  We can obtain the gravitino equation from
this Lagrangian,
\begin{equation}
\not\!\!
D\psi_{\mu}+m\psi_{\mu}=\left(D_{\mu}-\frac{m}{2}\gamma_{\mu}\right)
\gamma^{\nu}\psi_{\nu} .
\end{equation}
In solving this gravitino equation of motion, we gauge away the
goldstino( unitary gauge ) and use plane-wave ansatz for the
spatial dependence of $\psi_{\mu}\sim {e}^{i \boldmath k \cdot
\boldmath x }$.  Moreover it is convenient to decompose the space
component of gravitino field into the transverse part
$\psi_{i}{}^{T} $ and trace parts $\theta \equiv \gamma^i \psi_i$
and $k_i \psi_i$ as
\begin{equation}
\psi_i = \psi_i^T + \left( P_\gamma \right)_i \, \theta +
\left(P_k\right)_i k_i \psi_i ,
\end{equation}
where
\begin{eqnarray}
\left( P_\gamma \right)_i &\equiv& \frac{1}{2} \left(\gamma^i -
\frac{1}{\vec{k}^2} \, k_i \left( k_j \, \gamma^j \right)\right),
\nonumber\\
\left(P_k \right)_i &\equiv& \frac{1}{2 \, \vec{k}^2} \left( 3 \,
k_i - \gamma_i \left( k_j \, \gamma^j \right) \right).
\end{eqnarray}

This leads to the the following succinct form of dynamical field
equations which describe the degree of freedom corresponding to
transverse and longitudinal components,
\begin{eqnarray}
\left[ \gamma^0 \, \partial_0 + i \, \gamma^i \, k_i +
\frac{\dot{a} \, \gamma^0}{2} + \frac{a \, \underline{m}}{M_p^2}
\right] \psi_{i}{}^{T} = 0\, , \label{eq:t} \\ \left(\partial_0 +
\hat{B} + i \, \gamma^i \, k_i \, \gamma^0 \, \hat{A} \right)
\theta - \frac{4}{\alpha \, a} \, k^2 \, \Upsilon = 0\,,
\label{eq:l}
\end{eqnarray}
where
\begin{eqnarray}
\Upsilon = K_j{}^i \left( \chi_i \, \partial_0 \, \phi^j + \chi^j
\,
\partial_0
\, \phi_i \right)\nonumber\\ \underline{m} = P_R \, m + P_L \,
m^*\,,\qquad
\vert m \vert^2 = \underline{m}^\dag \, \underline{m}\, \nonumber\\
\hat{A} = \frac{1}{\alpha} \left( \alpha_1 - \gamma^0 \, \alpha_2
\right)
\nonumber \\
\hat{B} = - \frac{3}{2} \, \dot{a} \, \hat{A} + \frac{1}{2 \,
M_p^2} \, a \,
\underline{m} \, \gamma^0 \left(1 + 3 \, \hat{A} \right) \nonumber \\
\alpha =
3 \, M_p^2\left(H^2+ \frac{\vert m \vert^2}{M_p^4} \right)\nonumber\\
\alpha_1 = - M_p^2 \left(3 \, H^2 + 2 \, \dot{H} \right) -
\frac{3}{M_p^2} \, \vert m \vert^2, ~\alpha_2 = 2 \dot{
\underline{m}^\dag }. \label{manydef}
\end{eqnarray}
We can easily see, reducing the equation into this form,
eqn(\ref{eq:t}) describing the transverse component of gravitino
$\psi_i^T$ is decoupled from the longitudinal gravitino component,
and its coupling to the other fields are Planck mass suppressed.
So we hereafter pay our attention to the equation which describes
the longitudinal component of gravitino, eqn(\ref{eq:l}).
The form of $\Upsilon$ in (\ref{manydef}) tells us that, in the
absence of K\"ahler terms which mix the various left chiral
superfields, we need only worry about the fermionic partners of
dynamical scalar fields.  Furthermore, for our superpotential,
there is no mixing between the fermion associated with $\phi_1$
and those of $\phi_2$ and $\phi_3$, as long as $\phi_2=\phi_3=0$
which is true because they stay at the origin due to R-symmetry
once they roll down to the origin during the inflation.  Thus,
even though the effective masses of the fermions corresponding to
$\phi_2$ and $\phi_3$ are changing, those fermions do not
contribute to the goldstino and we can
 concentrate on the evolution of the other fields for the purpose
 of our calculation.

Based on the form of equation of motion involving two chiral
superfields,
 we can infer the terms in the Lagrangian which describe the
interactions of the
 two fields of our interests, namely, $ \theta$ ( longitudinal
component of
 gravitino) and $\Upsilon$ ( combination of chiral fermions orthogonal
to
 goldstino $\upsilon$).
Those interaction terms lead to the equation of motion
 in the following matrix form,

\begin{equation}
\left( \gamma^0 \, \partial_0 + i \, \gamma^i \, k_i N + M \right)
X = 0\,,\label{eqn:fermioneom}
\end{equation}
 with vector $X \equiv\left (\frac {\tilde \theta} {\tilde \Upsilon}
\right)$ consisting of canonically normalized fields
\begin{eqnarray}
\theta &=& \frac{2 \, i \, \gamma^i \, k_i}{\left(\alpha \, a^3
\right)^{1/2}} \, {\tilde \theta}\,, \nonumber\\ \Upsilon &=&
\frac{\Delta}{2} \left( \frac{\alpha}{a} \right)^{1/2} {\tilde
\Upsilon} ,
\end{eqnarray}
and diagonal mass matrix $M$ is given by
\begin{eqnarray}
M = \mbox{diag} \Big( ~~~ \frac{m a}{2 M_p^2} + \frac{3}{2} \left(
\frac{m a}{
M_p^2 } \alpha_1 + \dot{a} \alpha_2 \right), \nonumber \\
-\frac{ma}{2M_p^2}+\frac{3}{2}\frac{ma}{M_p^2}
\tilde{\alpha_1}+\dot{a} \tilde{\alpha_2}+a(m_{11}+m_{22}) ~~~
\Big)
\label{eqn:massmatrix}
\end{eqnarray}
and matrix $N$
\begin{equation}
N
\equiv \pmatrix{- \tilde{\alpha_1} & 0 \cr 0 & - \tilde{\alpha_1}
} + \gamma^0 \pmatrix{- \tilde{\alpha_2} & - \Delta \cr - \Delta &
\tilde{\alpha_2} } \label{kaptilde}
\end{equation}
for $ {\tilde \alpha_i} \equiv \alpha_i / \alpha$ and
$\Delta=\sqrt{1-{\tilde \alpha_1}^2-{\tilde \alpha_2}^2}$.  The
existence of off-diagonal terms warns us the non-trivial mixing of
fermion eigenstates.

Once we can reduce the Lagrangian into this form of matrix
expression, we can
 obtain the evolution equations for the mode functions ( the function
 multiplying the creation/annihilation operator) $U_r^{ij}$ and
$V_r^{ij}$ (
 i,j runs over 1 and 2 for two field case and $r$ for helicity $\pm$)
and
 calculate the occupation number of gravitino created from vacuum
through these
 mode functions.  In general, one is interested in the physical mass
 eigenstates
 ($\psi_1,\psi_2 $) which are non-trivial combinations (
with matrix coefficients) of gravitino $ \theta$ and matter chiral
fermions $\Upsilon$. So, strictly speaking, one would need to
diagonalize the Hamiltonian at each moment of field evolution to
keep track of the mass eigenstates and their abundance. This
diagonalization process is rather involved\footnote{ We refer the
readers to \cite{marco} for the general discussion.}. Here we use
a further simplification for our numerical analysis because we
are only interested in the asymptotic value of these abundances.
Indeed, since the mixing is small at such late times, we can simply
follow the fields of interest ($\theta,\Upsilon$).  That is, at
late times, these fields are approximate mass eigenstates and,
therefore, their occupation numbers correspond to those of
($\psi_1,\psi_2 $). The validity of this approximation will be
confirmed if we find that our occupation numbers cease to evolve
at the time scales of interest.
        
With this simplification, we may represent the mode decomposition in
the following familiar form,
\begin{eqnarray}
\lefteqn{ X^i \left( x \right) = \nonumber} \\ & & \int \frac{d^3
\mathbf{k}}{\left( 2 \pi \right)^{3/2}} e^{i \mathbf{k \cdot x}}
\Big[ U_r^{ij} \left( k,\, \eta \right) a_j^r \left( k \right) +
V_r^{ij} \left( k,\, \eta
\right) b_j^{\dagger r} \left( - k \right) \Big]. \nonumber \\
\end{eqnarray}
We then define the spinor matrix $U_{-}$ and $U_{+}$
\begin{equation}
U_r^{ij} \equiv \left[ \frac{U_+^{ij}}{\sqrt{2}} \, \psi_r,
\frac{U_-^{ij}}{\sqrt{2}} \, \psi_r \right]^T,\qquad V_r^{ij}
\equiv \left[ \frac{V_+^{ij}}{\sqrt{2}} \, \psi_r,
\frac{V_-^{ij}}{\sqrt{2}} \, \psi_r \right]^T \label{spin}
\end{equation}
with eigenvectors of the helicity operator $\mathbf{\sigma}\cdot
\mathbf{v} / \vert \mathbf{v} \vert$, $\psi_{+} =$
{\scriptsize{$\pmatrix{1 \cr 0}$}} and $\psi_{-} =$
{\scriptsize{$\pmatrix{0 \cr 1}$}}.  Using these spinor matrices,
the field equation of motion (\ref{eqn:fermioneom}) has a
following simple form in terms of the matrices $U_+$ and $U_-$,
\begin{equation}
a(t)\dot{U_\pm} = - i \, k \, U_\mp \mp i \, M \, U_\pm .
\label{numerical}
\end{equation}

We can then expand $U_\pm$ in terms of positive and negative
frequency solutions,
\begin{eqnarray}
U_+(t) \equiv \left( 1 + \frac{M}{\omega} \right)^{1/2}
\mathrm{e}^{- i \int^t \omega \, d t'} \, A \nonumber \\ -\left( 1
- \frac{M}{\omega} \right)^{1/2} \mathrm{e}^{i \int^t \omega \,
dt'} \, B \nonumber\\ \equiv \left( 1 + \frac{M}{\omega}
\right)^{1/2} \alpha - \left( 1 - \frac{M}{\omega} \right)^{1/2}
\beta\,, \nonumber\\ U_-(t) \equiv \left(1 - \frac{M}{\omega}
\right)^{1/2} \mathrm{e}^{- i \int^t \omega \, dt'} \, A \nonumber \\
+\left( 1 + \frac{M}{\omega} \right)^{1/2} \mathrm{e}^{i \int^t
\omega \, dt'} \, B\nonumber\\ \equiv \left( 1 - \frac{M}{\omega}
\right)^{1/2} \alpha + \left( 1 + \frac{M}{\omega} \right)^{1/2}
\beta \,\label{decoferm},
\end{eqnarray}
where diagonal matrix $\omega \equiv \sqrt{k^2+M^2} $.  $\alpha$
and $\beta$
  are precisely the generalization of Bogolubov coefficients.  Indeed,
in the
  same way as Bogolubov coefficients, we can calculate the
occupation
  number of i$^{th}$ fermion eigenstates from $\beta$ as
\begin{equation}
N_i \left( t \right) = \left( \beta^* \beta^T \right)_{ii} \mbox{
( no summation for $i$ )}. \label{numferm}
\end{equation}
We also keep in our mind that, because of nontrivial mixing of
fermion mass eigenstates for the case of coupled field system, we
need an extra care about the identification of inflatinos and
gravitinos.

\begin{figure}[h]
\centerline{\psfig{file=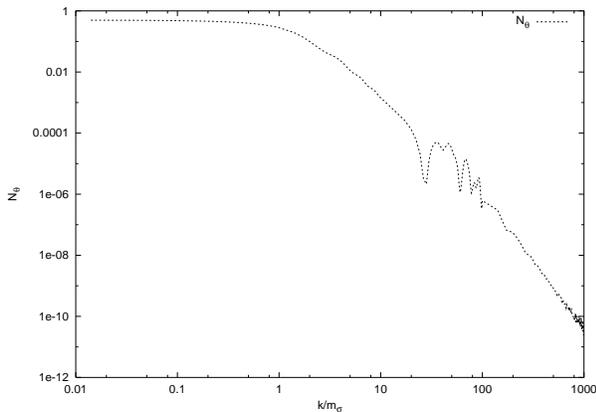,width=0.45 \textwidth}} \caption{
Gravitino abundance as a function of comoving momentum in units of
inflaton mass. } \label{fig1}
\end{figure}

We solved the coupled mode equations~(\ref{numerical}) numerically to
obtain the occupation numbers for $N_\theta$ and $N_\upsilon$.
These are plotted in Figure~(\ref{fig1}) as a function of comoving momentum
at time 1000 in  units of inflaton mass $m_{\sigma} \sim
\frac{\Delta^2}{M_p}$ which gives a typical time
 scale for inflaton oscillations.  We have used the typical parameter
values $\hat{\mu}=0.0001, \hat{\lambda}=0.001$ with an initial
inflaton amplitude $0.2 M_p$.  The O'Raifeartaigh field $\Phi_1$
has an initial amplitude $\hat{\mu}^2 M_p $ and we normalized
$a(t)$ to be one at the start of our calculation.

$N_{\theta}$ in our plot corresponds to the field abundance whose
mass converges asymptotically to gravitino mass $\vert m \vert /M_p^2 $ which
shows up in the first element of the mass matrix $M$ in
eqn(\ref{eqn:massmatrix}), and it gives an estimate
of gravitino abundances produced in preheating era.
We point out that, in
the presence of time dependent background as usually the case in
dealing with cosmological problems, kinetic term of
scalar fields can cause the supersymmetry breaking as 
we can easily see from the
supersymmetry transformation of chiral fermion, $f_{\chi}$(
superpartner of scalar $\chi$, say),
\begin{equation}
\delta f_{\chi}=
-\frac{1}{2}P_L\left[m_{\chi}-\frac{\gamma^0}{\sqrt{2}}\frac{d\chi}{dt}
\right]\epsilon ~~, \label{eqn:breaking}
\end{equation}
where $m_{\chi}$ is defined in eqn(\ref{eqn:m}).
Therefore we should be aware that the values for $N_1$ and
$N_2$ at intermediate time ( i.e. the time when $\sigma$ and
$\phi_1$ are still far from its settlement in the vacuum) does not
represent either of the inflatino or gravitino abundance 
because there exist
non-negligible contributions of supersymmetry breaking from both
inflaton and hidden sectors.
The time when this SUSY breaking contribution of $\dot{\sigma}
 $ and $m_{\sigma}$ becomes comparable with that of
 supersymmetry breaking sector is beyond the range of our numerical
integration. 
We checked, however, that in such a small parameter range 
with so small initial amplitude for $\phi_1$ as in our model, 
 $N_{1}$ and $N_{2}$ converge to asymptotic values and do
 not change anymore at relatively early
stage even when $\sigma$ still keeps its oscillation.
This verifies that Fig.\ref{fig1} should represent a good asymptotic behavior 
for gravitino abundance.
Especially, the cut-off scale $(k \sim m_{\sigma})$ of the
comoving momentum for the number density does not change anymore.
This reflects the fact that the oscillation scale of high momentum
mode $(k \gtrsim m_{\sigma})$ is much larger than that of the
background field $(m_{3/2})$, so high momentum modes behave
adiabatically and do not result in the non-adiabatic
amplification anymore in pre-heating era.

We further comment on the subtle problems in 
identification of gravitinos and inflatinos in terms of
mass eigenstates. Whatever field
which dominates the local supersymmetry breaking 
is considered to be longitudinal components of 
gravitinos via super Higgs mechanism.  
The 
non-thermal production of fermionic fields is efficient
 just after inflation when the kinetic term of inflaton 
field governs the supersymmetry breaking in our model. 
This is nothing but the non-adiabatic field amplifications of 
longitudinal components of gravitinos 
which 'eats' the fermionic partner of inflaton field, i.e. 
{\it{inflatino}}.
Hence when the fermion preheating is robust just
after inflation, it is to-be inflatino, not
 to-be gravitino of our interests, that is amplified via 
parametric resonance effects as
 longitudinal components of gravitinos free from 
Planck suppression.

%

We also should mention that, because of the couplings, hidden
sector may not be the sole cause for supersymmetry breaking even
in the vacuum, and in fact, there could be still supersymmetry
breaking from inflaton sector in the vacuum at a later time. It,
however, can be shown that in the vacuum the supersymmetry
contribution from inflaton sector is at most of order $\vert F
\vert^4$ compared with $\vert F \vert^ 2 ~( \sim \mu ^4 ) $ due to
hidden sector \cite{hitoshi,weinberg}, and we still observe the
dominant contribution of local supersymmetry breaking from the
O'Raifeartaigh field at a later time. We thus expect our plot of
$N_{\theta}$ still represents a fairly good overall behavior of
gravitino abundance in asymptotic regime.

For the comparison with the gravitino number density constraints
from photo-dissociation process in nucleosynthesis for the case of
thermal production of gravitinos in thermal bath (\ref{eqn:ns}),
we need to integrate $N_{\theta}(k)$ over the comoving momentum
space.  As usually the case with the preheating of fermions, our
plot also indicates that occupation number as a function of
comoving momentum k can be as large as of order unity at most up
to the order of inflaton mass scale, $k_{cutoff} \simeq m_{\sigma}
$ and decreases exponentially for bigger k.  So the number density
for longitudinal components,
\begin{equation}
n_{3/2}=\frac{1}{\pi^2}\frac{1}{a^3}\int_0^{k_{max}} \vert \beta_k
\vert ^2 k^2 dk
\end{equation}
during the preheating is at most
\begin{equation}
n_{3/2}(t_{pre}) \lesssim k_{cutoff}^3 \simeq m_{\sigma}^3 \simeq
\frac{\Delta^6}{M_p^3}\simeq 10^{-25} M_p{}^3. \label{eqn:npre}
\end{equation}
We can now estimate the upper bound of the ratio of gravitino
number density $n_{3/2}$ to entropy density in analogy with (
\ref{eqn:orns} ).
\begin{equation}
 \frac{n_{3/2} (t_r)}{s(t_r)} \simeq\frac{n_{3/2}(t_{pre})\left(
\frac{a(t_{pre})}{a(t_r)} \right )^3}{0.44 g_* T_{RH}^3} \simeq
\frac{n_{3/2}(t_{pre})\left( \frac{t_{pre}}{t_r} \right )^2}{0.44
g_* T_{RH}^3} ,
\end{equation}
and substitution of $1.66 \cdot T_{RH}^2 \sqrt{g_*} / M_p \sim H
\sim t_r^{-1}$ for $t_r$ gives us the estimate of $n_{3/2}/s$
after reheating,
\begin{equation}
\frac{n_{3/2}}{s} \simeq \frac{n_{3/2}(t_{pre}) t_{pre}^2 T_{RH}(
1.66)^2} {0.44 M_p^2} .
\label{eqn:gravitinos}
\end{equation}
We expect this efficient gravitino production occurs well within
the time range of typical oscillation of supersymmetry breaking
field and we can substitute $t_{pre} \sim 1/m_{\phi_1} \sim
(\alpha_{\kappa} \mu)^{-1} \sim 10^9 M_p^{-1}$ and eqn
(\ref{eqn:npre}) in above equation to obtain the upper bound,
\begin{equation}
\frac{n_{3/2}}{s} \lesssim 6.3 \cdot 10^{-15}
\left(\frac{T_{RH}}{10^{10}\mbox{GeV}}\right).
\end{equation}
This upper bound of $n/s$ for the gravitinos from nonthermal
production in our model is thus smaller than eqn(\ref{eqn:ns}) of
thermal scattering by at least two orders of magnitude.

We therefore find that our model does not lead to the
overproduction of gravitinos due to nonthermal process in
preheating period, and reheating temperature constraint due to
this effect is less severe than that of gravitinos produced by the
scattering in thermal bath during the reheating period. We also
point out that the expression given by eqn(\ref{eqn:gravitinos})
was derived in a general setting and it can be used to obtain, in
combination with eqn(\ref{eqn:ns}), the estimate for the relative
significance of the gravitino production in thermal and
non-thermal processes once the model of supergravity inflation is
given.

\section{Conclusion and Discussion}

We showed in this letter a realistic supergravity inflation model
which breaks local supersymmetry in the vacuum dominantly via
$F$-term coming from O'Raifeartaigh field in the hidden sector.
We emphasized the significance of radiative correction in
supersymmetry breaking sector to evade the moduli problem, and
subsequently obtained the non-minimal K\"ahler potential arising
from this loop correction. Using this non-minimal form of K\"ahler
potential, we analyzed the possible non-thermal production of
gravitinos in preheating era.
The emphasis in our analysis was on the longitudinal components of 
gravitinos which do not suffer from Planck suppression and 
hence potentially could lead to robust amplification 
via parametric resonance effects. 
We showed that the comoving number density of fermionic 
mass eigenstate
 converges to its asymptotic value at an early stage of 
preheating era which is
well before the time when the supersymmetry contribution 
comes from 
hidden sector fields.
Physically, this indicates that the large comoving modes
($ k \gtrsim m_{\sigma}$) is much larger than the typical coherent oscillation
of background fields at later times ($k \sim m_{3/2}$ ).
Hence comoving number density for big modes behave adiabatically
 and we do not expect the parametric amplifications 
at later times.
We also discussed the subtle problems in 
identification of gravitinos and inflatinos due
to the  non-trivial mixing 
of fermionic mass eigenstates.
To-be {\it{inflatinos}} are longitudinal components of 
{\it{gravitinos}} just after inflation, and its role
 is replaced by fermionic partner of O'Raifeartaigh fields
at later times.  

We estimated the upper bound of number density of non-thermally
produced gravitinos by integrating out its comoving occupation
number over momentum space. Because of the small mass scale of
inflaton field, the typical momentum scale of produced gravitinos
and consequently the momentum space over which occupation number
is integrated out turns out to be small as well. This leads to the
relatively small number density of gravitinos and we showed that
it gives less significant constraint than that of gravitinos which
are produced by thermal scattering.

We point out that the cases involving three and more superchiral
fields are rather involved. We can basically follow the formalism
discussed in section \ref{sec:nonthermal}, but we need additional
care in interpreting the fermion fields as a superposition of mass
eigenstates because we cannot completely gauge away one of fermion
fields via unitary gauge as we can do in the two field case
\cite{long,marco}.  The case including the gauge interaction terms
and the model of other supersymmetry breaking mechanism besides
hidden sector supersymmetry breaking are also to be examined.

\acknowledgments 
We wish to thank L.~Kofman for helpful comments and insights 
and S.~Sarkar for providing us with useful references. We also 
thank J.~Cohn, M.~Peloso and A.~Pierce for fruitful discussions.
H.M. was supported by NSF under grant PHY-0098840 and DOE contract
DE-AC03-76SF00098. PBG was supported by the DOE and the NASA grant
NAG 5-10842 at Fermilab.

\end{document}